\documentclass[prl,preprint,showpacs,preprintnumbers]{revtex4}
\usepackage{colordvi}
\usepackage{amssymb}
\usepackage{amsmath}
\usepackage{epsf}
\usepackage{mathrsfs} 
\usepackage[dvips]{graphics}
\usepackage{graphicx, subfigure,psfrag}
\usepackage[dvips,usenames]{color}
\begin{document}
\def \beq{\begin{equation}}
\def \eeq{\end{equation}}
\def \bea{\begin{eqnarray}}
\def \eea{\end{eqnarray}}
\def \bem{\begin{displaymath}}
\def \eem{\end{displaymath}}
\def \P{\Psi}
\def \Pd{|\Psi(\boldsymbol{r})|}
\def \Pds{|\Psi^{\ast}(\boldsymbol{r})|}
\def \Po{\overline{\Psi}}
\def \bs{\boldsymbol}
\def \bl{\bar{\boldsymbol{l}}}
\title{Reversal of Klein reflection in bilayer graphene} 
\author{Neetu Agrawal (Garg)$^1$, Sameer Grover$^{2}$,  Sankalpa  Ghosh$^2$ and  Manish Sharma$^1$}
\affiliation{$^1$Centre for Applied Research in Electronics, Indian Institute of Technology Delhi, New Delhi-110016, India}
\affiliation{$^2$Department of Physics, Indian Institute of Technology Delhi, New Delhi-110016, India}
\begin{abstract}
Whereas massless Dirac fermions in monolayer graphene exhibit Klein tunneling when passing through  a potential barrier  upon normal incidence,  such a barrier totally reflects massive Dirac fermions in bilayer graphene due to difference in chirality. We show that, in the presence  of magnetic barriers, such massive Dirac fermions can have transmission through  even at normal incidence. The general consequence of this behaviour for multilayer graphene consisting of massless and massive modes are mentioned. We also briefly discuss the effect of a bias voltage on such magnetotransport.
\end{abstract}
\pacs{73.43.-f,81.05.Tp,72.90.+y,73.23.-b,73.63.-b,78.20.Ci,42.25.Gy}{}
\date{\today}
\maketitle

One of the most remarkable features of electron transport in graphene \cite {Castroneto} is that charge carriers are  chiral in nature  \cite{katnovo}
and their degree of chirality changes as the number of layers  are varied \cite{Geimbilayer, Falkobilayer, MinMac}. In  monolayer graphene, chirality  results in  Klein tunneling, namely the perfect transmission of a normally incident electron of energy E through a potential barrier of height V when  $E < V$.  In contrast to this, for  bilayer graphene chirality of a different degree leads to total reflection from the same barrier at normal incidence, a phenomenon often called Klein reflection \cite{KSN3,sim1}. Thus, chirality strongly influences coherent ballistic  transmission through graphene based heterostructures, a phenomenon which is now experimentally accessible  \cite{Levitovmasir, Kim1, gordon, YoungKim} and potent with  possibilities for new device applications. 


For monolayer graphene, it has been shown that the combined effect of an inhomogenous magnetic field, dubbed as a magnetic barrier,  and an electrostatic potential significantly alters ballistic transport of charge carriers. It has already been shown  that use of such inhomogenous magnetic barriers can lead to confinement of  chiral electrons as opposed to Klein tunneling \cite{MDE07}. This has subsequently led to a large body of work \cite{magbarrier}.

In this Letter, we consider the effect of such barriers in the case of bilayer graphene. Fabry-Perot like transmission fringes develop due to  scattering of chiral charge carriers from inhomogeneous magnetic barriers in the presence of voltage. By analyzing these fringes, we find that, in contrast to the above discussion on monolayer graphene, it is quite the opposite in bilayer graphene. It is seen that it is possible to actually reverse the Klein reflection of normally incident electrons; namely, the barrier can cause transmission and significantly modifies the observed conductance in bilayer graphene.  As we show, this also paves the way of a more generalized understanding of the effect of such magnetic barriers on transport of chiral quasiparticles in multilayer graphene which shares many features of monolayer and bilayer graphene.

\begin{figure}
\begin{center}
\centerline{\epsfxsize 5.5cm \epsffile{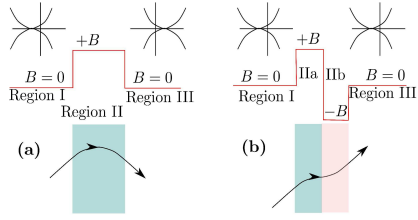}
}
\end{center}
\caption{\textit{E-k diagram and electron trajectory in bilayer graphene with (a) magnetic barrier, and (b) barrier + well }}
\label{schematic}
\end{figure}

The generalized  chiral operator of degree $J$ that describes chirality of  charge carriers in multiple layer graphene with $ABAB\cdots$ type of stacking is  $ \hat{h}_{\bs{p}}^{J}=\bs{n}_{J} \cdot \sigma$, where  
$\bs{n}_{J} = \cos (J \phi) \hat{x} + \sin (J \phi) \hat{y}$, $\phi = \tan^{-1}\frac{p_y}{p_x}$ and $ \sigma$ is the two-component Pauli matrix.   Monolayer and bilayer graphene  are just two cases of this description with $J=1$ and  $J=2$ respectively. The general  Hamiltonian for  charged quasiparticles of $J$-th degree chirality is then  
\beq H_{J} \propto |\bs{p}|^{J} h_{\bs{p}}^{J}  \label{chiral}  \eeq 
 The above Hamiltonian leads to a  pseudo Zeeman effect due to a momentum-dependent pseudo magnetic field $\bs{B_p}=|p|^{J} \bs{n}$ \cite{GM}. The corresponding eigenstates with definite chirality of degree $J$ are given as 
 \beq | + \rangle = \frac{1}{\sqrt{2}}\begin{bmatrix} e^{ -i \frac{J\phi}{2}} \\ e^{i \frac{J\phi}{2}} \end{bmatrix} ;~ |- \rangle =  \frac{1}{\sqrt{2}} \begin{bmatrix} e^{ -i \frac{J\phi}{2}} \\- e^{i \frac{J\phi}{2}}
\end{bmatrix} \label{spinor} \eeq
These eigenvalues are proportional to $ \pm |p|^{J}$ for the conduction ($+$) and the valence ($-$) bands respectively and give both massless and massive Dirac modes for multilayer graphene in general \cite{MinMac}. Eq.(\ref{spinor}) also demonstrates how the pseudospin winding number described in Ref.\cite{park} varies with the number of layers. 

The upper and lower components of the pseudospinor give the probability amplitude of an electron being in sublattice $A$ and $B$. A pure  scalar potential  $V(\bs{r})$  cannot flip the pseudospin, namely $\langle \pm  | V(\bs{r}) | \mp \rangle=0$, since it couples in the same way with both sublattices. For monolayer graphene with $J=1$, there is only one massless
Dirac mode. Since a scalar potential shifts the Dirac point, to preserve pseudospin, an  electron (hole) state outside the potential barrier needs to be matched  with a hole(electron) state inside the barrier  and that leads to perfect transmission. 

The present work discusses  bilayer graphene in Bernal stacking, which corresponds to $J=2$  in Eq.(\ref{chiral}) and has one massive Dirac mode. Here also, a scalar potential cannot flip the pseudospin and shifts the Dirac point. However, now the spectrum is parabolic. Hence, a  state with $\bs{p}^2$  outside the potential barrier  will go to a state with  $-\bs{p}^2$  inside the barrier to preserve the pseudospin. This requires that a propagating wave outside the barrier should be matched with an evanescent wave inside the barrier leading to total reflection at  normal incidence.

Unlike a scalar potential, a vector potential due to a magnetic field $\bs{B}=B(x,y)\hat{z}$  couples with the momentum and can flip the pseudospin. In the Landau gauge the  vector potential is $\bs{A}=A_y(x) \hat{y}$, one can show that when such a magnetic field  is present the pseudospin projection angle changes from $J\phi$  to $J\phi'$ where $\phi' = \tan^{-1}( \frac{q_y}{q_x})$, where $q_y = k_y + \frac{e A_{y}}{{\hbar c}}$ (cf. Fig.\ref{schematic}(a)). 
Here, we use $\bs{k} = \{k_x, k_y \}$ with no magnetic barrier and $\bs{q}= \{q_{x}, q_y \}$
with a magnetic vector potential. The eigenstates are then
 \beq | + \rangle_{B} = \frac{1}{\sqrt{2}}\begin{bmatrix} e^{ -i \frac{J\phi'}{2}} \\ e^{i \frac{J\phi'}{2}} \end{bmatrix} ;~ |- \rangle_{B} =  \frac{1}{\sqrt{2}} \begin{bmatrix} e^{ -i \frac{J\phi'}{2}} \\- e^{i \frac{J\phi'}{2}}
\end{bmatrix}.  \label{magspinor} \eeq
Scattering between one of the eigenstates in Eq.(\ref{spinor}) with  $\bs{B} = 0$ to one in  Eq.(\ref{magspinor}) for $\bs{B}\neq 0$ is now possible as the pseudospin is not necessarily conserved in the presence of a magnetic field.  Consequently, perfect Klein reflection upon normal incidence will  not take place in bilayer graphene in presence of such a magnetic barrier.  We show this by explicitly calculating transmission by a transfer matrix approach \cite{masir, nilsson}. Chiral charge carriers in bilayer graphene obey 
\beq 
 i\hbar\frac{\partial \Psi (x,y)}{\partial t} = H\Psi,~~~
 H=\begin{bmatrix} V_1 & \Pi & t & 0 \\ \Pi^{+} &V_1 & 0 & 0 \\ t & 0 & V_2 & \Pi^{+} \\ 0 & 0 & \Pi & V_2 \end{bmatrix} \label{hammat} \eeq
Here,  they are described by a $4$-component spinor $\Psi(x,y)  =\begin{matrix} ( \Psi_a & \Psi_b & \Psi_c & \Psi_d)^T \end{matrix}$ and a $4 \times 4$  Hamiltonian in the presence of a magnetic barrier  and an electrostatic potential. $\Pi = v_F[p_x+i(p_y+eA/c)]$  with Fermi velocity $v_{F} =10^6$ m/s. The magnetic field profile is $\bs{B}=B \Theta(x^2 - d^2) \hat{z}$,  $ V_1$ and $V_2$ are the potentials at the two different layers, and $t$ is the tunnel coupling between the two layers. This Hamiltonian reduces to a manifestly
chiral symmetric form given in Eq.(\ref{chiral}) in the limit $ \frac{|E|}{t} \ll 1$, where $E$ is the incident electron energy \cite{Falkobilayer}. $V_{1}= (\neq)V_{2}$  correspond to unbiased (biased) graphene bilayer cases. The stationary solutions $\Psi= \psi(x,y)e^{-\frac{iEt}{\hbar}}$ which obey $H \psi =E\psi$ can be explicitly obtained  in the three regions $ x < -\frac{d}{2}$ (region I), $-\frac{d}{2} \le x \le \frac{d}{2}$ (region II) and $x > \frac{d}{2}$ (region III). In regions I ($j=1$) and III ($j=2$) these solutions are propagating as well as evanescent waves. In region II where the magnetic field is finite, stationary solutions are given by  parabolic cylindrical functions. We define $\ell_B=\sqrt{\frac{\hbar c}{eB}}$ and  $\epsilon_B=\frac{\hbar v_{F}}{\ell_B}$  as units of length and energy respectively and  introduce dimensionless variables  $x \rightarrow \frac{x}{\ell_B}$ , $v_{1,2}=\frac{V_{1,2}}{\epsilon_B} $ and $\epsilon =\frac{E}{\epsilon_B}$. 
In regions I and III, the dispersion relations now become 
\beq [-q(x)^2 + (\epsilon^{'}+\delta)^2][-q(x)^2 + (\epsilon^{'}-\delta)^2]= ({\epsilon^{'}}^2-\delta^2){t^{'}}^2  \nonumber \eeq

Here, $q(x)^2 = q_{x}^2 + q_{y}^2$, $q_{y}(x) = k_y + sgn(x) \pi \frac{\Phi}{\Phi_0}$, $\epsilon^{'} = \epsilon-v_0$ with  $v_0=(v_1+v_2)/2=\frac{V_{+}}{\epsilon_B}$ and $\delta = (v_1-v_2)/2 = \frac{\Delta}{\epsilon_B}$.  Also, $\frac{\Phi}{\Phi_0}$ is the total magnetic flux through an area $d \ell_B$ in units of the flux quantum $\Phi_0=\frac{hc}{e}$, and $\ell_B=\frac{\hbar c}{eB}$ is the magnetic length.  For an evanescent wave, $q_{x}^2 =-\kappa_x^2 < 0$.

We now begin with the case of an unbiased graphene bilayer where $\delta=0$ and $v_{1}=v_{2}=v_{0}$. In regions I and III we set $v_{1,2}=v_{0}=0$. The transmission through such a combination of magnetic barrier and electrostatic barrier can be written using transfer matrices $ \mathcal{M}_{0}(x)$ and $\mathcal{M}_{B}(x)$, where the subscript $0$ ($B$) defines the region 
with magnetic field $0$ ($B$). These matrices are given below explicitly.
\begin{widetext} 
\bea \mathcal{M}_{B}(x) & =  & \begin{bmatrix} D_{p^{+}}(z) & D_{p^{-}}(z) & D_{p^{+}}(-z) & D_{p^{-}}(-z) \\
 \varepsilon_{1}^{*}p^{+}D_{p^{+}-1}( z) & \varepsilon_{1}^{*}p^{-} D_{p^{-}-1}( z) & \varepsilon_{1}p^{+}D_{p^{+}-1}( -z) &  \varepsilon_{1}p^{-}D_{p^{-}-1}( -z) \\ \varepsilon_{2}^{+}D_{p^{+}}( z) & \varepsilon_{2}^{-}D_{p^{-}}( z) & \varepsilon_{2}^{+}D_{p^{+}}( -z)& \varepsilon_{2}^{-}D_{p^{-}}( -z)\\ \
\varepsilon_{1}\varepsilon_{2}^{+}D_{p^{+}+1}(z)& \varepsilon_{1}\varepsilon_{2}^{-}
D_{p^{-}+1}(z)& \varepsilon_{1}^{*}\varepsilon_{2}^{+}D_{p^{+}+1}(-z)&
\varepsilon_{1}^{*}\varepsilon_{2}^{-}D_{p^{-}+1}(-z) \end{bmatrix}  \nonumber \\
\mathcal{M}_{0}(x) & = & \frac{1}{\epsilon'}\left[ \begin{matrix} {\epsilon'}e^{i q_x x} & {\epsilon'}e^{-i q_x x} & {\epsilon'}e^{-\kappa_x x} & {\epsilon'}e^{ \kappa_x x}  \\  
[{q_x-iq_{y}(x)}]e^{iq_x x} & -[{q_x+iq_{y}(x)}]e^{-i q_x x} &  i[{ \kappa_x-q_{y}(x)}]e^{- \kappa_x x}    &     -i[{\kappa_x+q_{y}(x)}]e^{\kappa_x x} \\ 
-{\epsilon'}e^{i q_x x} &-{\epsilon'} e^{-i q_x x} & {\epsilon'}e^{-\kappa_x x} & {\epsilon'}e^{ \kappa_x x}     \\  
- [{q_x+iq_{y}(x)}]e^{iq_x x} & [{q_x-iq_{y}(x)}]e^{-i q_x x} &  i[{ \kappa_x+q_{y}(x)}]e^{- \kappa_x x}    &     -i[{\kappa_x-q_{y}(x)}]e^{\kappa_x x}    \end{matrix} \right] \label{transfer}
 \eea
\end{widetext}
Here, $z=\sqrt{2}(x + k_{y}), p^{\pm} = (\gamma_{\pm}-1)/2, \varepsilon_{1}=\frac{i \sqrt{2}}{\epsilon^{'}}$ and $\varepsilon_{2}^{\pm}= \frac{\epsilon^{'}}{t'}-\frac{2 p^{\pm}}{t'\epsilon^{'}}$, and $\gamma_{\pm} = \epsilon'^2 \pm \sqrt{1+\epsilon'^2 t'^2}$.
The current density expression is obtained as $j_x = v_F\psi^{+}\left(\begin{matrix} \sigma_x & 0 \\ 0 & \sigma_x \end{matrix}\right)\psi $.   
The transfer matrix through any combination of a scalar and vector potential  
can now be written in terms of  transfer matrices given in Eq.(\ref{transfer}) for regions with finite and zero magnetic field.  This can then be used to find the transmission. The ratio of current density in region III and the incident current density in region I gives the transmission probability as a function of the angle of incidence $\phi=\tan^{-1}\frac{k_y}{k_x}$. 

\begin{figure}
\begin{center}
\centerline{\epsfxsize 3.6cm \epsffile{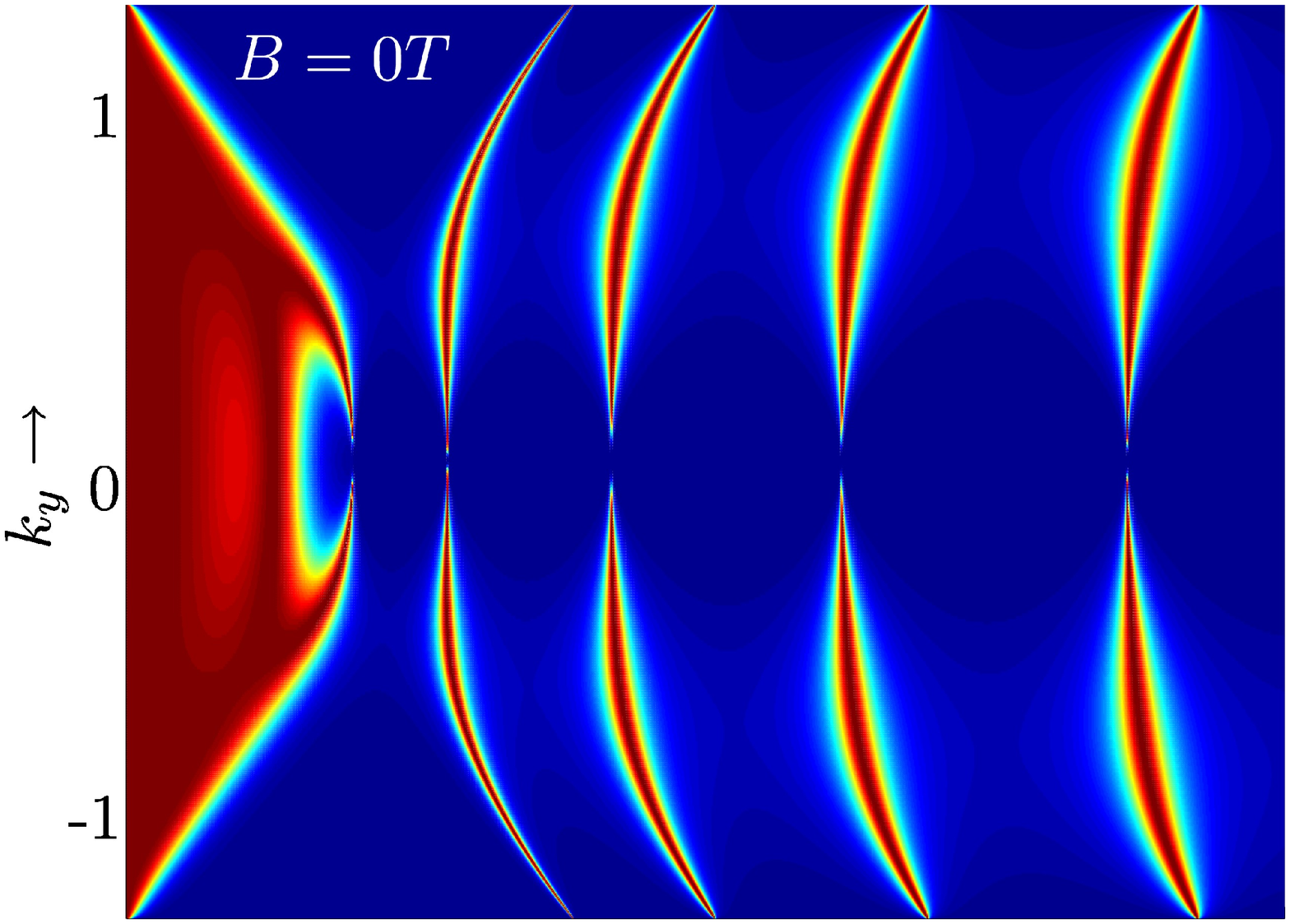}\epsfxsize 3.6cm \epsffile{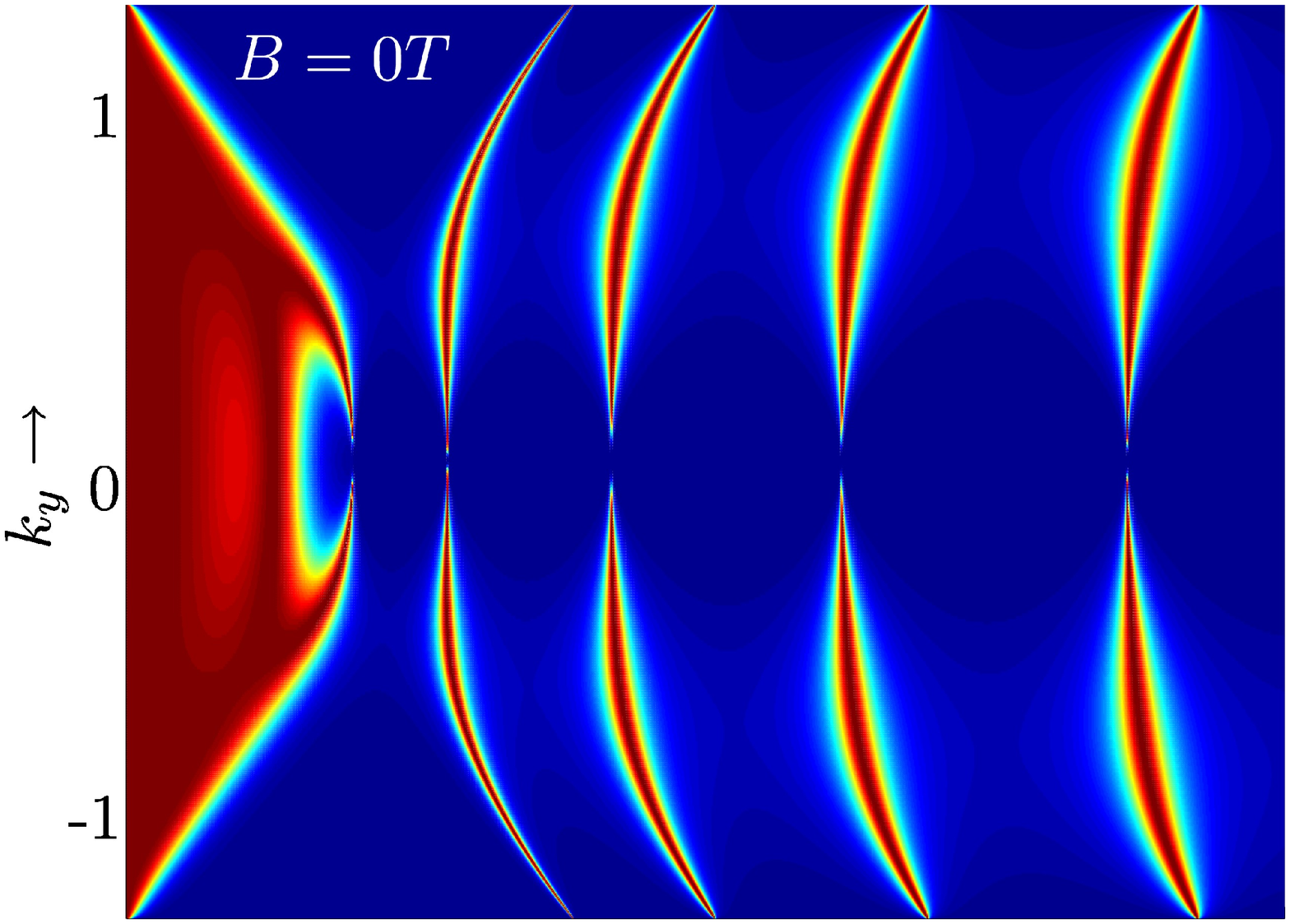}}
\centerline{\epsfxsize 3.67cm \epsffile{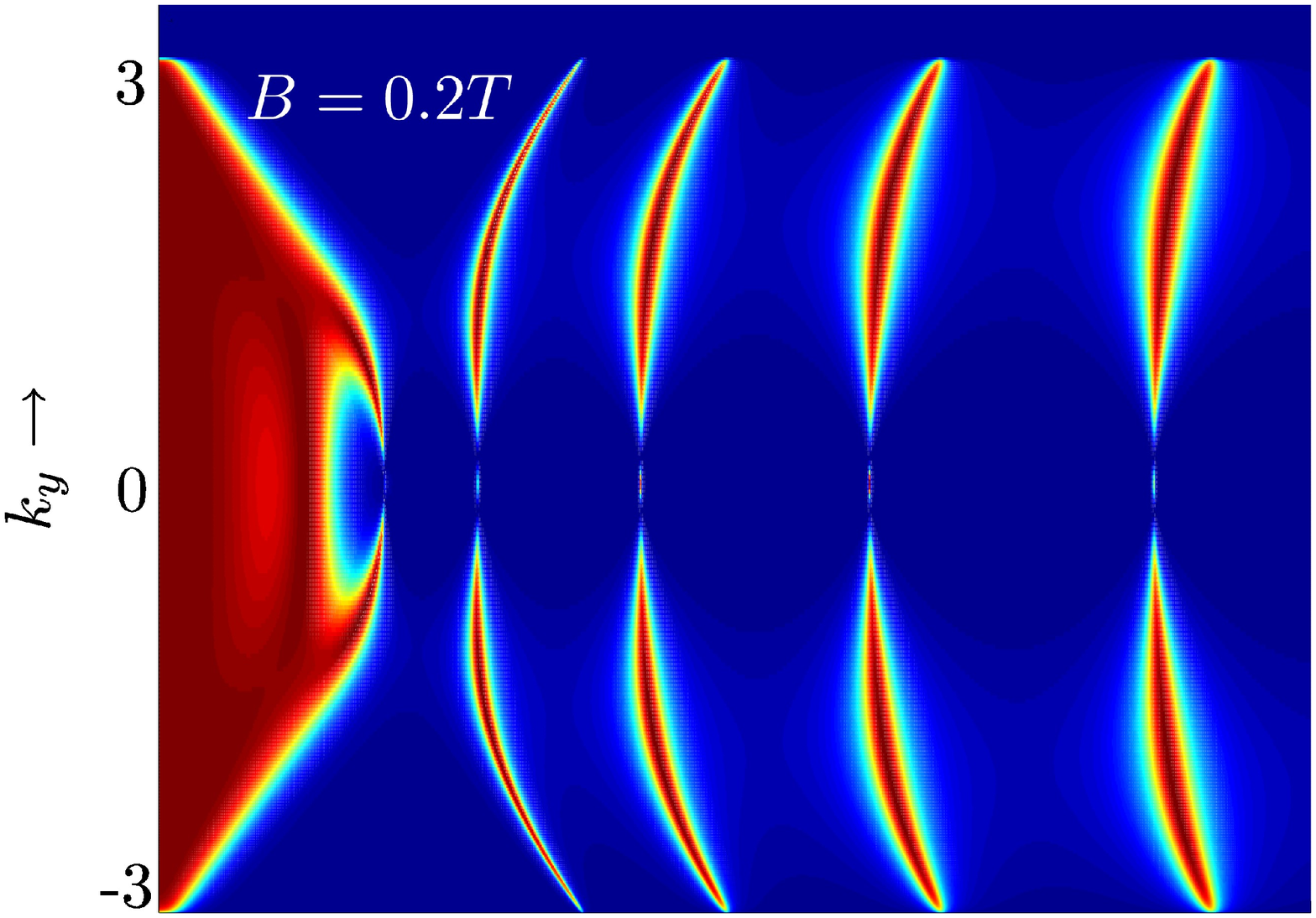}\epsfxsize 3.6cm \epsffile{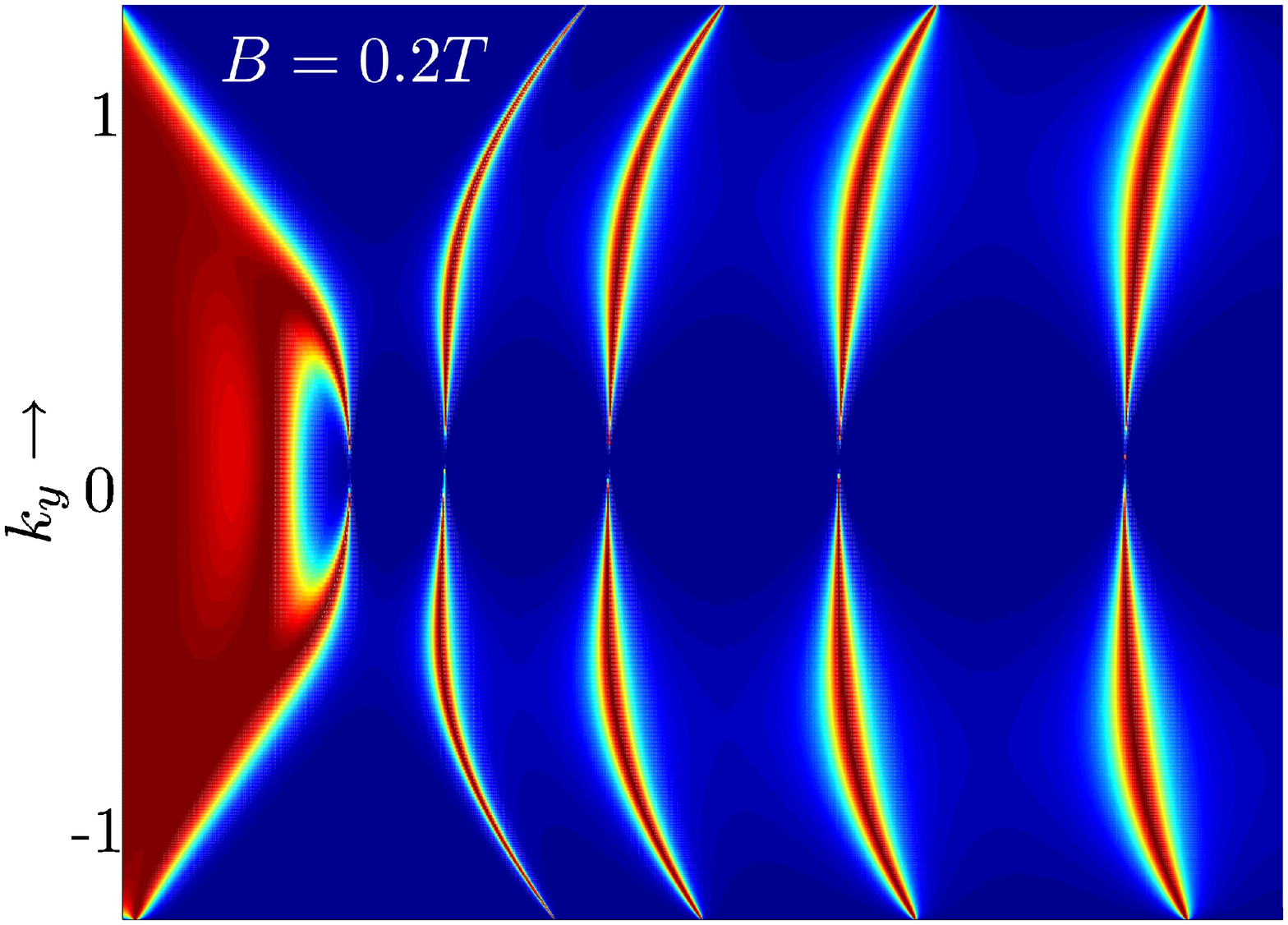}}
\centerline{\epsfxsize 3.67cm \epsffile{2d.eps}\epsfxsize 3.6cm \epsffile{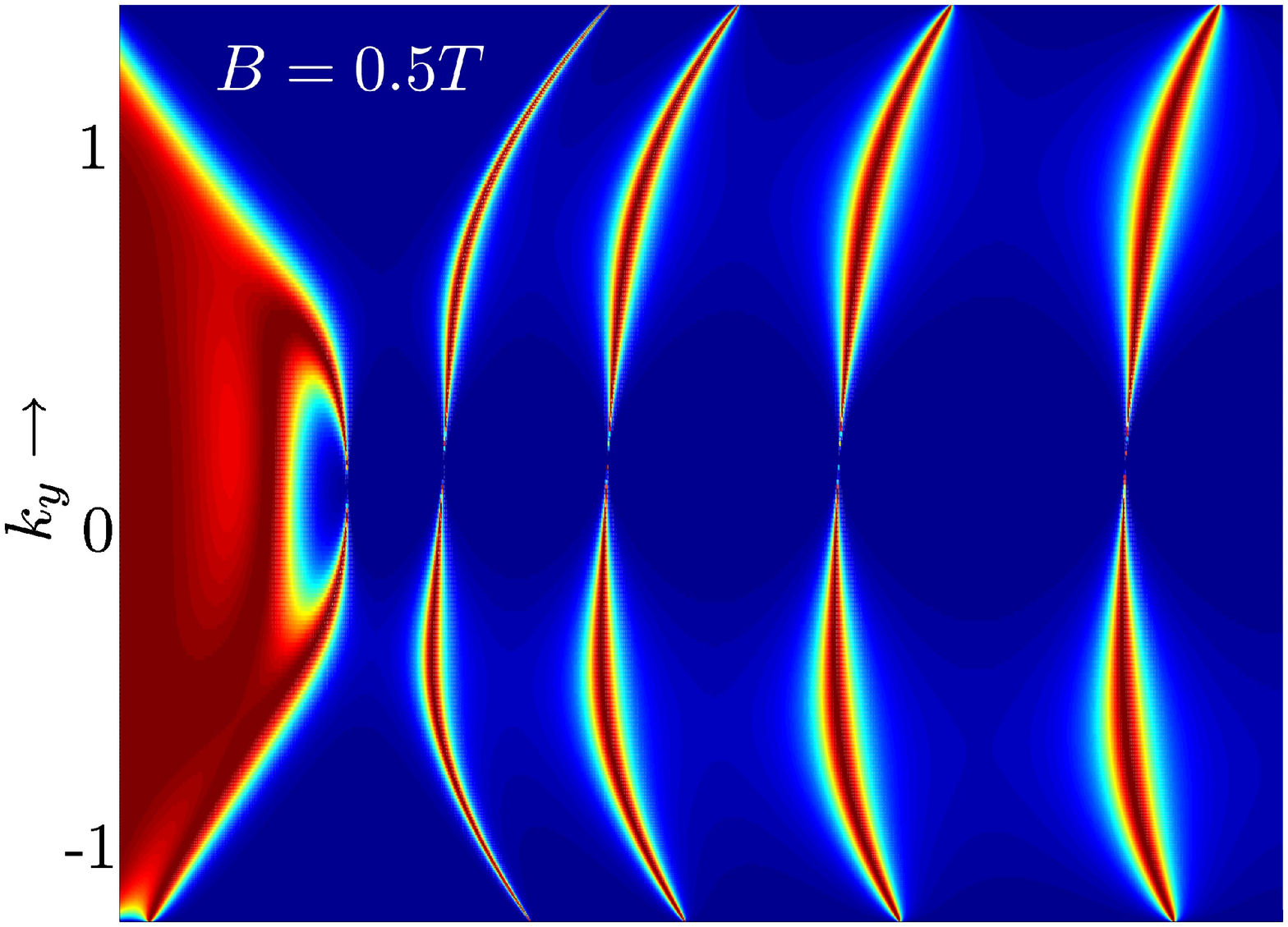}}
\centerline{\epsfxsize 3.75cm \epsffile{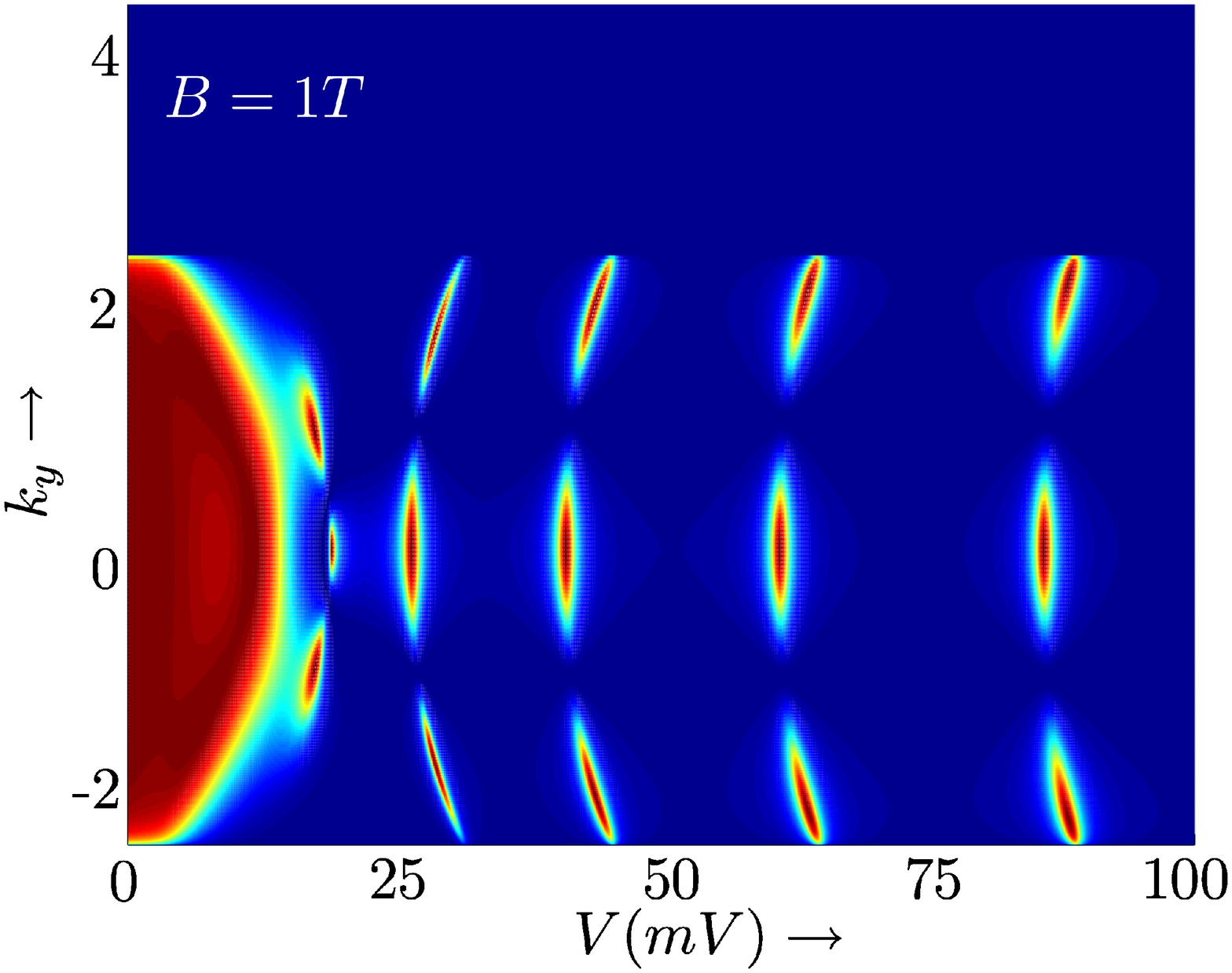}\epsfxsize  3.73cm \epsffile{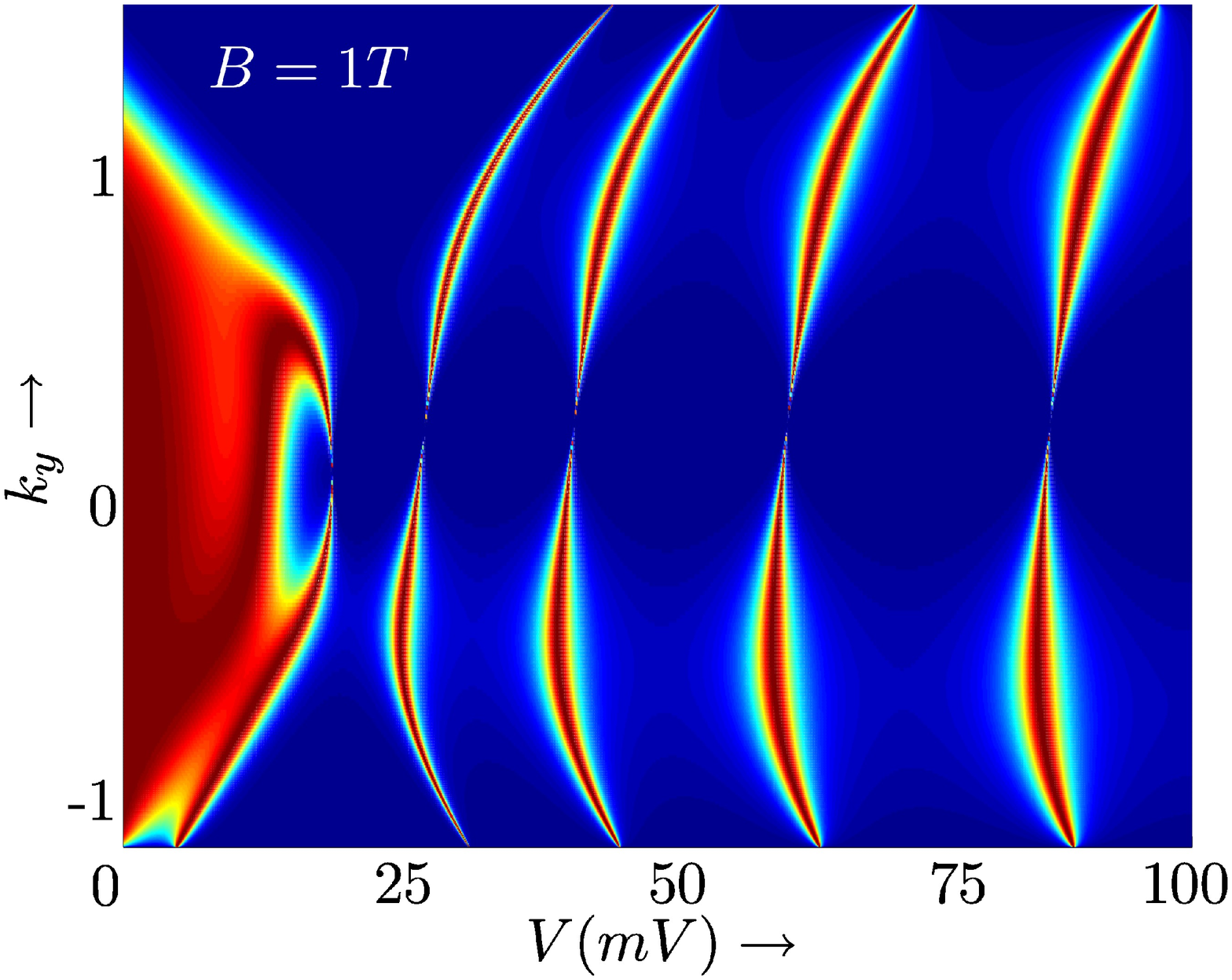}}
\end{center}
\caption{\textit{Transmission T as a function of  incidence angle for different fields. (left) Barrier, (right) Barrier+well.}}
\label{caseone}
\end{figure}

In Fig.\ref{caseone}, the plots in the left column show transmission through a magnetic barrier with  increasing strength as a function of the incidence angle and the applied voltage in the barrier regime. The uppermost figure shows the transmission for $B=0$, which clearly shows the region of  perfect reflection at and around the normal incidence symmetrically placed between  two wings of resonant Fabry-Perot fringes. This is a generic feature of transmission through such barriers \cite{Levitovmasir}. As the strength of the barrier increases, a transmission region develops between these two wings due to the effect of magnetic barrier and the resulting transmission also becomes asymmetric. Clearly seen is the disappearance of perfect Klein reflection at normal incidence. The total angular range of transmission however shrinks in the presence of the magnetic barrier since, beyond a certain incident angle, all electron waves suffer total internal reflection. This is an important result of our paper since it shows the exact reason why an inhomogeneous barrier causes Klein reflection to be reversed to  transmission for bilayer graphene. 

In the right column in Fig.\ref{caseone} is plotted similar transmission through a setup consisting of two magnetic barriers, equal in magnitude and opposite in direction such that the total flux through the region vanishes, as shown in Fig.\ref{schematic}. Now, the incident and the transmitted wave vectors are parallel to each other. As a result, the net rotation of the pseudospinor due to the inhomogenous field vanishes and Klein reflection at normal incidence is restored. The bending of the Fabry-Perot fringes can be attributed to the asymmetry in the angular transmission at a given voltage in presence of  magnetic barrier $+$ well combination.

\begin{figure}
\begin{center}
\centerline{\epsfxsize 3.8cm \epsffile{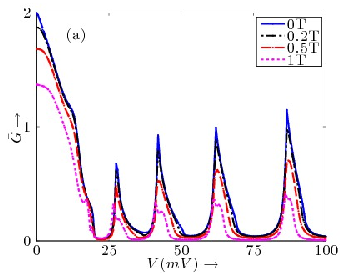}\epsfxsize 3.8cm \epsffile{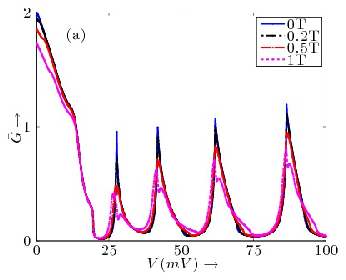}}
\end{center}
\caption{\textit{${\bar{G}}$ for  different magnetic barrier strengths. (a)  Barrier,  (b) Barrier+well. $\Delta=0$,  $E=17meV$ kept constant.
}}
\label{barrierwell}
\end{figure}

We now study the effect of the above transmission on conductance at very low temperatures and for energies close to the Fermi energy in the linear transport regime. To a good approximation, the dimensionless conductance $\bar{G}$ can be written as \cite{magbarrier}
\beq \bar{G}=\int_{-\frac{\pi}{2}}^{\frac{\pi}{2}} d \phi T(E, \phi) \cos \phi  \eeq
A comparison between the conductance with and without a magnetic barrier in Fig.\ref{barrierwell} shows that the gaps between the conductance maxima and minima get reduced in the presence of a magnetic barrier. The barrier reflects electrons incident upon it beyond a critical angle  and the angular range of transmission shrinks with increasing barrier strength. As a result, the absolute value of conductance maxima comes down. For the barrier$+$well case, the conductance is slightly higher. This detailed characterization of ballistic transport through  magnetic barriers dictated by pseudospin chirality is one of the main results of this work. 

The preceding discussion shows how the effect of  localized magnetic field  on transport of massive Dirac modes encountered in bilayer graphene is very different as compared to that of massless Dirac modes in monolayer graphene. Another important observation can be made by looking at magnetotransport through magnetic barriers for monolayer graphene \cite{MDE07} and the present analysis for bilayer graphene. The same barrier which is reflective for the massless mode is transmissive for the massive mode. The general theory of  N-layer graphene predicts the existence of $1$ massless and $N-1$ massive modes for $N$ odd and only massive modes for $N$ even \cite{MinMac}. The magnetic barrier thus can be used to at least partially filter out or to selectively allow a mode.

\begin{figure}
\begin{center}
\centerline{\epsfxsize 3.9cm \epsffile{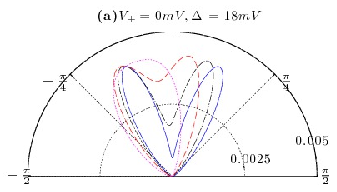}\epsfxsize 3.9cm \epsffile{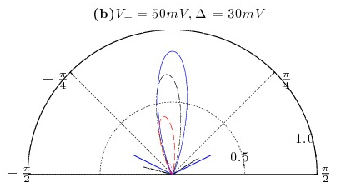}}
\centerline{\epsfxsize 3.9cm \epsffile{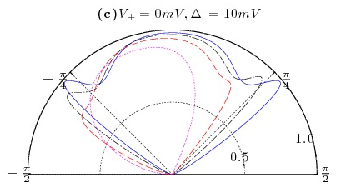}\epsfxsize 3.9cm \epsffile{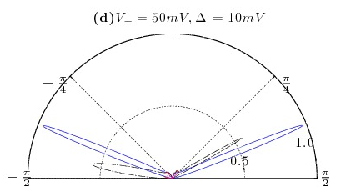}}
\end{center}
\caption{\textit{Biased bilayer transmission for  various $V_+$, $\Delta$ and magnetic field strengths (colours and lines same as in Fig.\ref{barrierwell}).}}
\label{figure3}
\end{figure}

Finally, we  briefly discuss the influence of magnetic barriers on transport through a biased bilayer; i.e., when 
$ V_{1} \neq V_{2}$ in Eq.(\ref{hammat}). The effective $ 2 \times 2$ Hamiltonian is 
\bea H_b & = &  \begin{bmatrix} V_{1}  &  -\frac{v_{F}^2}{t} ( \Pi^{\dagger})^2  \\
                                            -\frac{v_{F}^2}{t} (\Pi)^2   & V_{2}  \end{bmatrix} \nonumber \\
 & = & H_{ub} + \Delta \sigma_z + V_{+}\mathbf{1}  \label{bbl} \eea
$H_{ub}$ is the Hamiltonian for the unbiased  bilayer given in Eq.(\ref{chiral}) for $J=2$. Thus,  bias introduces an additional $z$ component of the pseudo magnetic field. This will try to rotate the pseudospin out of the $x-y$ plane. If $\Delta$ is large  and the incident energy lies in this 
gap, the pseudo Zeeman gap will suppress transmission. This is because pseudospin-flipping is energetically costly, a phenomenon that has been experimentally observed \cite{biasedbl1, biasedbl2, bandgap1, bandgap2}. The third term in Eq.(\ref{bbl}) does not flip the pseudospin.  
Combining the first two terms,  the Hamiltonian can be rewritten as
\bea  H_b  =  ( \bs{n} \cdot \bs{\sigma}){\sqrt{ \Delta^2 + \left(\frac{\hbar^2 q^2 v_F^2}{t}\right)^2}} &  &    \nonumber   \eea 
Here, $\bs{\sigma}$ are three-component Pauli matrices and $\bs{n}  = ( \sin \theta \cos  2\phi', \sin \theta \sin 2\phi',  \cos \theta )$ with $\tan\theta =  \frac{\hbar^2 q^2 v_{F}^2}{t \Delta}$. The corresponding pseudospinors $|  + \rangle = 
(\cos \frac{\theta}{2} e^{ -i {\phi'}}, \sin \frac{\theta}{2} e^{ i \phi'} )^{T}$ and 
 $|  - \rangle = 
(\sin  \frac{\theta}{2} e^{ -i {\phi'}}, -\cos \frac{\theta}{2} e^{ i {\phi'}} )^{T}$ 
 are again eigenstates with definite chirality. Comparing these states with the ones given in Eq.(\ref{spinor}), we find that  the magnetic barrier gives an in-plane rotation while the bias voltage  gives an out-of-plane rotation to the pseudospinor. The two orthogonal rotations can also be given in different spatial regions and have been recently discussed using a Berry phase argument \cite{sim2} for a pure bias. In this paper, we have considered the case where $\Delta$ and $\bf B$  are non-zero  in the same region. If the gap is large and the incident energy lies inside the gap (Fig.\ref{figure3}(a)), the transmission  as expected  
 is highly suppressed and there is no visible effect of the magnetic barrier. Also, we find that, as long as the incident energy does not lie in the gap (Fig.\ref{figure3}(b), (d)) or the gap is too small (Fig.\ref{figure3}(c)), the effect of the  barrier on the biased and the unbiased bilayers is similar.

This work is supported by grant SR/S2/CMP-0024/2009 of DST, Govt. of India. One of the authors (NA) acknowledges support from CSIR, Govt. of India.

\end{document}